\documentclass[nofootinbib,showpacs,preprintnumbers,amsmath,amssymb,floatfix]
{revtex4-1}
\usepackage{epsf}

\begin{document}


\title{New look at the QCD factorization}

\vspace*{0.3 cm}

\author{B.I.~Ermolaev}
\affiliation{Ioffe Physico-Technical Institute, 194021
 St.Petersburg, Russia}
  \author{M.~Greco}
\affiliation{Department of Physics and INFN, University Roma Tre,
Rome, Italy}
\author{S.I.~Troyan}
\affiliation{St.Petersburg Institute of Nuclear Physics, 188300
Gatchina, Russia}

\begin{abstract}
We show that both the $k_T$- and collinear factorization for DIS structure functions can be
obtained by consecutive reductions of the Compton scattering amplitude. Each of these reductions
is an approximation valid under certain assumptions. In particular, the transitions to the
$k_T$- factorization is possible when the virtualities of the partons connecting the perturbative
and non-perturbative blobs are space-like. Then, if the parton distribution
has a sharp maximum in $k_{\perp}$, the $k_T$ factorization can be reduced to the
collinear factorization.
\end{abstract}


\maketitle

\section{Introduction}
The QCD factorization is the fundamental concept to provide theoretical grounds for applying the
Perturbative QCD to description of hadronic reactions. According to the factorization, any
scattering amplitude $A$ in QCD can be represented as a convolution of a perturbative (E) and
non-perturbative (T) contributions:
\begin{equation}\label{factgen}
A = E \otimes T
\end{equation}
There are two kinds of the factorization
in the literature: Collinear factorization\cite{colfact} and the $k_T$- factorization\cite{ktfact}
where the DIS structure functions $f(x,Q^2)$
are respectively represented as follows:
\begin{equation}
\label{colfact}
f(x,Q^2) = \int_x^1 \frac{d \beta}{\beta} f^{(pert)}(x/\beta,Q^2/\mu^2) \phi (\beta, \mu^2)
\end{equation}
and
\begin{equation}
\label{ktfact}
f(x,Q^2) = \int_x^1 \frac{d \beta}{\beta} \int
\frac{d k^2_{\perp}}{k^2_{\perp}} f^{(pert)}(x/\beta,Q^2/k^2_{\perp}) \Phi (\beta, k^2_{\perp})
\end{equation}
where $f^{(pert)}$ stand for the perturbative components of the structure functions;
$\phi$ and $\Phi$ are the parton distributions and $\mu$ is the factorization scale.
In what follows we obtain Eqs.~(\ref{colfact},\ref{ktfact}), simplifying the factorized expression for
the amplitude $A_{\mu\nu}$ of the Compton scattering off a hadron target. By doing so, we
summarize and generalize the results obtained in \cite{egtfact}.  Using
appropriate projection operators $P_r$ the Compton amplitude $A_{\mu\nu}$
can be expanded into a set of invariant amplitudes $A_r$. According to the Optical Theorem,
every structure function $f_r$ can be expressed through $A_r$:

\begin{equation}\label{opt}
f_r = \frac{1}{\pi} \Im A_r
\end{equation}
Among amplitudes $A_r$ there is the amplitude
$A_S$ related to the structure function $F_1$ singlet. We will address this amplitude as
the singlet and will address as non-singlets to all other invariant
amplitudes and use for them the generic notation $A_{NS}$. We also will use the generic notation
$A$ for both the singlet and non-singlet amplitudes when it is relevant.

\section{Basic Factorization for the Compton amplitude}

\begin{figure}[h]
\begin{center}
\begin{picture}(400,180)
\put(0,0){ \epsfbox{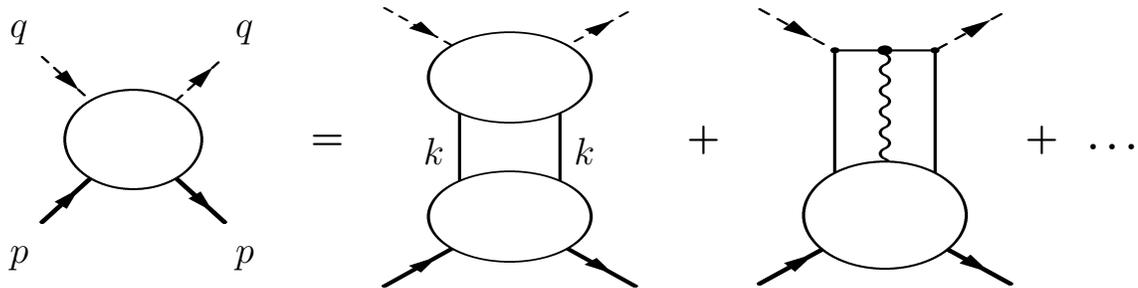} }
\end{picture}
\end{center}
\caption{\label {dspinfig1} Representation of $A_{\mu\nu}$ through the convolution of
two blobs.}
\end{figure}

Let us expand the invariant amplitude $A$ into a set of convolutions depicted in Fig.~
\ref{dspinfig1}
where the $t$- channel  states involve arbitrary number of partons.

Throughout the paper
we will consider only
the first graph in Fig.~1 where the blobs are connected by the two-parton state,
with the partons being quarks. Consideration of the two-gluon state yields the
same results as shown in \cite{egtfact}. All blobs
in Fig.~1 can contain both perturbative and non-perturbative contribution, so this
kind of factorization does not correspond to the conventional scenario of the QCD
factorization. We will address it as the primary convolution. Introducing the
Sudakov parametrization of the moment $r$:
\begin{equation}\label{sud}
k = -\alpha (q + x p) + \beta p + k_{\perp} ,
\end{equation}
 we can write  the primary convolution as follows, using the :

\begin{equation}\label{primfact}
A (q^2,w) = \int_{-\infty}^{\infty} \frac{d \beta}{\beta}
\int^{\infty}_0 d k^2_{\perp}
\int_{-\infty}^{\infty}d \alpha
\widetilde{A} (w \beta, q^2, k^2) \frac{B}{\left(k^2\right)^2}
T (w\alpha, k^2) ,
\end{equation}
where $\widetilde{A}$ and $T$ denote the upper and lower blobs respectively;
$w = 2pq$, $k^2 = - w \alpha \beta - k^2_{\perp}$ and factor $B$, with
$B = w(\alpha^2 + \beta^2) + k^2_{\perp}$, appears because of simplification of the spin
structure of the intermediate quarks. We have skipped in Eq.~(\ref{primfact})
dependence on unessential arguments like masses, spin, etc. The integrand
in Eq.~(\ref{primfact}) becomes singular at $k^2 \to 0$. This infrared ($\equiv$ IR) divergence
must be regulated. The IR-sensitive perturbative contents for the singlet and non-singlet amplitudes are
different. $A_{NS}$ contain the IR-sensitive perturbative logarithms whereas $A_S$ includes both
logarithms and the power-factor:

\begin{equation}\label{irterms}
A_{NS} = A_{NS} \left(\ln (w \beta), \ln (Q^2/k^2)\right),~~
A_{S} =  \left(w \beta/k^2\right) M_{S} \left(\ln (w \beta), \ln (Q^2/k^2)\right) .
\end{equation}
Therefore in order to keep the integral Eq.~(\ref{primfact}) IR stable ,
amplitudes $T$ must obey the following restrictions at small $k^2$:

\begin{equation}\label{tir}
T_{NS} \sim \left(k^2\right)^{\gamma},~~~T_{S} \sim \left(k^2\right)^{1+\gamma},
\end{equation}
with $\gamma > 0$.
Similarly, in order to get the ultraviolet stability of $A$
the blob $T$ at large $\alpha$ should decrease with growth of $|\alpha|$:

\begin{equation}\label{tuv}
T_{NS} \sim |\alpha|^{-1 - h}, ~~~~T_{S} \sim |\alpha|^{- h}.
\end{equation}

\begin{figure}[h]
\begin{center}
\begin{picture}(140,160)
\put(0,0){ \epsfbox{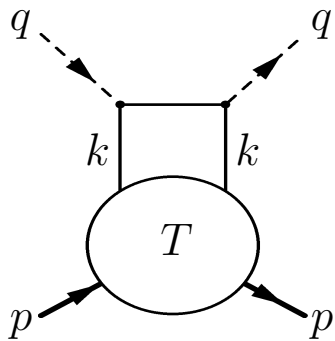} }
\end{picture}
\end{center}
\caption{\label {dspinfig2} Born approximation for the
amplitude of the forward Compton scattering.}
\end{figure}

Eq.~(\ref{primfact}) in the Born approximation
is depicted in Fig.~\ref{dspinfig2}.

\begin{figure}[h]
\begin{center}
\begin{picture}(280,140)
\put(0,0){ \epsfbox{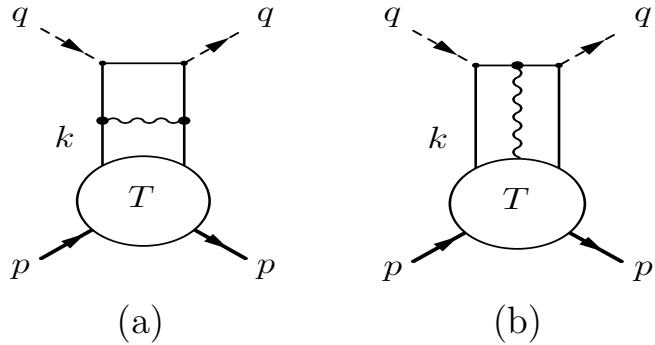} }
\end{picture}
\end{center}
\caption{\label {dspinfig3} Radiative corrections to the Born
amplitude.}
\end{figure}

Radiative corrections are absent there, so blob $T$ is totally
non-perturbative. Inserting the radiative corrections into the Born approximation is depicted in
Fig.~\ref{dspinfig3}.

We stress that we neglect graphs with extra propagators touching the lower blob (e.q. graph (b))
because they lead to the convolution with three or mote intermediate partons depicted in Fig.~1
and we do not consider such multiparton states in this paper. In order to back up
this course of actions we would like to notice that all
evolution equations available operate with the two-parton initial states only.
So, we account for the graphs which do not touch it (e.q. graph (a)). Obviously all such graphs
can be included into the upper blob, leaving the lower blob non-perturbative. As a result, we
convert the convolution in Eq.~(\ref{primfact}) into the similarly looking convolution

\begin{equation}\label{basefact}
A (q^2,w) = \int_{-\infty}^{\infty} \frac{d \beta}{\beta}
\int^{\infty}_0 d k^2_{\perp}
\int_{-\infty}^{\infty}d \alpha
A^{(pert)} (w \beta, q^2, k^2) \frac{B}{\left(k^2\right)^2}
T (w\alpha, k^2) ,
\end{equation}
where the upper blob $A^{(pert)}$ is perturbative and the lower blob $T$ is non-perturbative.
The integral in  (\ref{basefact}) is free of IR singularities at small $k^2$. Therefore,
Eq.~(\ref{basefact}) corresponds to the concept of QCD factorization, though this factorization
differs from the collinear and $k_T$- factorizations. By this reasons we will address it as the basic
factorization. Applying Optical Theorem, we convert (\ref{basefact}) into the basic
factorization for the structure functions:

\begin{equation}\label{basicfact}
f (x, Q^2) = \int_{-\infty}^{\infty} \frac{d \beta}{\beta}
\int^{\infty}_0 d k^2_{\perp}
\int_{-\infty}^{\infty}d \alpha
f^{(pert)} (x /\beta, Q^2/ k^2) \frac{B}{\left(k^2\right)^2}
\Psi (w\alpha, k^2)
\end{equation}
where $\Psi$ stands for the totally unintegrated parton distributions.

\section{Reducing Basic factorization to $k_T$- and collinear factorizations}

In order to proceed from Eq.~(\ref{basicfact}) to (\ref{ktfact}), we need to integrate out the
$\alpha$- dependence without touching the perturbative . Obviously, it cannot be done
straightforwardly because $f^{(pert)} $ depends on $\alpha$ trough $k^2$. However, imposing the
restriction
\begin{equation}\label{k2}
w\alpha\beta \ll k^2_{\perp} ,
\end{equation}
we can neglect this dependence in and integrate $\Psi$ over $\alpha$. As a result we arrive at
 (\ref{ktfact}) with

\begin{equation}\label{phikt}
\Phi (\beta,k_{\perp} ) = \int_{k^2_{\perp}/w}^{k^2_{\perp}/w \beta}
d \alpha T (\alpha, k^2) .
\end{equation}

In order to keep (\ref{ktfact}) IR stable at $k_{\perp} \to 0$,
 the parton distributions $\Phi$ should decrease with $k_{\perp}$:

\begin{equation}\label{phiir}
\Phi_{NS} \sim \left(k^2_{\perp}\right)^{\gamma},~~~\Phi_{S} \sim \left(k^2_{\perp}\right)^{1+\gamma}.
\end{equation}

\begin{figure}[h]
\begin{center}
\begin{picture}(240,100)
\put(0,0){\epsfxsize=50mm \epsfbox{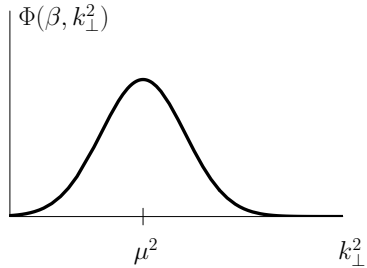} }
\end{picture}
\end{center}
\caption{\label {dspinfig4} The peaked form of $\Phi(\beta,k_{\perp}^2)$ with one maximum}.

\end{figure}

Transition from the $k_T$- expression (\ref{ktfact}) to the collinear factorization (\ref{colfact})
is also impossible in the straightforward way. Let us suppose that the $k_{\perp}$-dependence
of  $\Phi_{S,NS}$ in (\ref{ktfact}) has a peaked form with one or several sharp maximums.
 at $k^2_{\perp} = \mu^2_ 0, \mu^2_ 1,...$
as shown in Fig.~\ref{dspinfig4}. We address such scales as intrinsic scales.

We do not
assume any special form for the curve in Fig.~3 save that it obeys the restriction (\ref{phiir}).
It allows us to approximately integrate over $k_{\perp}$ in (\ref{ktfact}), dealing with $\Phi$ only
and arriving at

\begin{equation}
\label{colfactzero}
f(x,Q^2) = \int_x^1 \frac{d \beta}{\beta} f^{(pert)}(x/\beta,Q^2/\mu^2_0) \varphi (\beta, \mu^2_0)
\end{equation}
where the parton distributions $\varphi$ are expressed through the distributions $\Phi$
which have been used in the $k_T$- factorization:

\begin{equation}\label{phimu}
\varphi (\beta, \mu^2_0) = \int_0^w \frac{d k^2_{\perp}}{k^2_{\perp}}\Phi(\beta, k^2_{\perp}) .
\end{equation}

\section{Comparison of conventional collinear factorization and Eq.~(\ref{phimu}).}

The parton distribution $\phi$ in the conventional approach to the
collinear factorization and distribution $\varphi$ are widely different. The
distribution $\phi$ includes both perturbative and non-perturbative contributions
whereas $\varphi$ is purely non-perturbative. The factorization scale $\mu$ used in the
conventional approach is arbitrary while $\mu_0$ corresponds to the maximum in Fig.~4.
However, it is easy to relate them, using any kind of the perturbative evolution to
evolve $\varphi$ from scale $\mu_0$ to $\mu$. Naturally, the value of $\mu$ can be chosen
anywhere between $\mu^2_0$ and $Q^2$. At the same time the perturbative part,
$f^{(pert)}(x/\beta,Q^2/\mu^2_0)$, should be evolved from $\mu_0$ to $\mu$. As a
result, we arrive at the conventional formula (\ref{colfact}) where the
convolution is independent of $\mu$. In other words,
changing the factorization scale from
the intrinsic scale $\mu_0$ to an arbitrary scale $\mu$ leads to the re-distribution of
the radiative corrections between the upper and lower blobs of the collinear convolution.
We do not specify which kind of the perturbative evolution should be used
because our approach is insensitive to to details of this evolution.
In particular, he DGLAP equations can be used for such evolution.

\section{Restrictions on the DGLAP fits for the parton distributions}

Combining Eqs.~(\ref{tuv}, \ref{phikt}. \ref{phimu}) leads to the following dependence
of the parton distributions $\Phi$ and $\varphi$ at small $\beta$:

\begin{equation}\label{phibeta}
\Phi_{NS} \sim \beta^{h},~~\Phi_{S} \sim \beta^{-1 +h},~~
\varphi_{NS} \sim \beta^{h},~~\varphi_{S} \sim \beta^{-1 + h}.
\end{equation}

As shown in Eq.~(\ref{fit}), the standard DGLAP -fits for the DIS structure
functions in the collinear factorization
include
a normalization $N$, the singular factors $x^{-a}$, with $a > 0$,
and the regular terms:

\begin{equation}\label{fit}
\delta q, \delta g = N x^{-a} (1 - x)^b (1 + c x^d)\,,
\end{equation}
where the parameters $N, a, b, c, d > 0$.
Such expressions do not do not look as the ones
obtained with the perturbative methods, so we
identify them with non-pertrurbative distributions $\varphi$. Eq.~(\ref{phibeta})
excludes the use of the singular factors in the expressions for the non-singlet
structure functions $F_2, F^{NS}_1, g_1$, etc
and also suppress the singular factors with $a > 1$ in the
expressions for the singlet $F_1$. However, the parton  distributions used for
$F_1$ and $F_2$ are identical,
therefore the suppression of the singular factors with $a > 0$ can be
applied to all structure functions, including the singlet $F_1$. The singular
factors $x^{-a}$ in the DGLAP fits for initial parton densities should
be removed from the fits because they contradict to the
integrability of the basic convolutions of the Compton amplitudes.

\section{Conclusion}

Both the $k_T$- and collinear factorizations are obtained by consecutive reductions
of the Compton scattering amplitude represented as the convolution of two blobs
connected by two parton lines. We neglect all convolutions with
number of the intermediate states greater than two. It has no impact on
our further analysis because every convolution should be finite independently
of the multiplicity of intermediate states. Exploiting the IR stability of
the convolution we convert it into the basic QCD convolution and to the $K_T$ -
factorization. This transition is performed with purely mathematical means.
 In contrast, the transition from the $K_T$-to the collinear
factorization is based on the physical assumption:
we assume that the $k_{\perp}$- dependence of the parton distribution has one
or several sharp
maximums which become the intrinsic factorization scales.
The sharper the maximums are, the more accurate this reduction is. In order to keep
the lower blob unperturbative, the value of the intrinsic scale(s) should be close to $\Lambda_{QCD}$.
Our assumption of the peaked $k_{\perp}$- distributions can be checked by
analysis of experimental data in the framework of the
$k_T$-factorization. Transition to the conventional parton distributions $\phi$ defined at
other factorization scales $\mu$ located in the domain of the perturbative QCD
(conventionally $\mu \sim $ several GeV),
can be done with the use of the evolution equations. On the other hand, the perturbative
scale can be regarded as the one achieved with the perturbative evolution starting from
a lower scale which can be associated with our intrinsic scale $\mu_0$. Therefore,
the conventional approach involves the intrinsic scale, though implicitly, while our approach
sets this scale explicitly.

\section{Acknowledgement}

We are grateful to Organizing Committee of the workshop DSPIN-2001 for support.
The work is partly supported by Grant RAS 9C237,
Russian State Grant for Scientific School
RSGSS-65751.2010.2 and EU Marie-Curie Research Training
Network under contract MRTN-CT-2006-035505 (HEPTOOLS).

\end{document}